\documentclass[12pt,a4]{article}
\usepackage{amsmath,amssymb}
\usepackage{epsfig}

\newcommand{\im}{\mathrm i}

\newcommand{\tr}{\operatorname{Tr}}
\newcommand{\llg}{\operatorname{ln}}

\newcommand{\eq}{\begin{equation}}
\newcommand{\en}{\end{equation}}
\newcommand{\bear}{\begin{eqnarray}}
\newcommand{\ear}{\end{eqnarray}}
\newcommand{\bt} { \begin{tabular} }
\newcommand{\et}{ \end{tabular} }
\newcommand{\bc} { \begin{center} }
\newcommand{\ec}{ \end{center} }

\title{The thermal conductivity of alternating spin chains}

\author{G.A.P. Ribeiro\footnote{pavan@df.ufscar.br} \\ 
Universidade Federal de S\~ao Carlos, Departamento de F\'{i}sica, \\ 
CP 676, 13565-905 S\~ao Carlos-SP, Brazil \\ 
N. Cramp\'e\footnote{nicolas.crampe@lpta.univ-montp2.fr} \\ 
LPTA, UMR 5207 CNRS - Universit\'e Montpellier II \\ 
34095 Montpellier, France\\
A. Kl\"umper\footnote{kluemper@physik.uni-wuppertal.de} \\  Theoretische Physik, Bergische Universit\"at Wuppertal, \\ 
42097 Wuppertal, Germany}

\begin{document}
\maketitle
\begin{abstract}
We study a class of integrable alternating ($S_1,S_2$) quantum spin chains
with critical ground state properties. Our main result is the description of the thermal Drude weight of the
one-dimensional alternating spin chain as a function of temperature. We have
identified the thermal current of the model with alternating spins as one of
the conserved currents underlying the integrability. This allows for the
derivation of a finite set of non-linear integral equations for the thermal
conductivity. Numerical solutions to the integral equations are presented for specific cases
of the spins $S_1$ and $S_2$. In the low-temperature limit a universal picture
evolves where the thermal Drude weight is proportional to temperature $T$ and
central charge $c$.
\end{abstract}
\centerline{PACS numbers: 05.50+q, 02.30.IK, 05.70Jk} 
\centerline{Keywords: Bethe Ansatz, Thermal Drude weight, alternating spin chain}
\thispagestyle{empty}

\newpage
\section{Introduction}

In recent years considerable progress has been achieved in the understanding
of transport phenomena in low-dimensional strongly correlated quantum systems
\cite{ZOTOS0}. Thermal transport in (quasi) one-dimensional systems has been
investigated on the theoretical as well as on the experimental side
\cite{ZOTOS}. The existence of anomalous heat transport, indicated by for
instance a non-zero Drude weight, has been established in particular for
integrable quantum systems.  The temperature dependence of the thermal
conductivities of spin-$\frac{1}{2}$ Heisenberg chain compounds was measured
and revealed anomalous transport properties \cite{EXP1,EXP2,EXP3}. The
experimental results point into the same direction as the theoretical findings
obtained by the Bethe ansatz technique \cite{SAKAI} for the spin-$\frac{1}{2}$
Heisenberg chain and the spin-$S$ Heisenberg chain \cite{RIBEIRO}.

Here we are interested in the computation of the thermal conductivities of the
general case of integrable alternating spin $(S_1,S_2)$ chains using Bethe
ansatz techniques. Our main result is the computation of the thermal Drude weight yielding a
finite value for finite temperatures implying ballistic thermal
transport. Specifically, we have obtained a finite set of non-linear integral
equations providing the thermal conductivity as function of temperature. These
equations are solved numerically for specific values of $S_1$ and $S_2$.

The paper is organized as follows. In section \ref{INTEGRA}, we outline the
basic ingredients of the Bethe ansatz techniques needed in this work. In
section \ref{THERMAL}, we discuss the derivation of the set of non-linear
integral equations for the description of the thermal Drude weight. Here, we
also present our numerical and analytical findings for the solution of the
non-linear integral equations. Our conclusions are given in
section \ref{CONCLUSION}.

\section{Integrability and conserved currents}\label{INTEGRA}

In the theory of integrable models the generating function of quantum
integrals of motion is an object playing the role of a transfer matrix of the
associated classical statistical model. The transfer matrix is the trace of an
ordered product of Boltzmann weights defined on the square
lattice. Specifically we can define the transfer matrix $T(\lambda)$ of a
rotational invariant classical vertex model associated with the alternating spin
chain ($S_1,S_2$) as a product of two ordinary transfer matrices
$t^{(S_i)}(\lambda)$
\bear
T(\lambda)&=&t^{(S_1)}(\lambda)t^{(S_2)}(\lambda), \label{ordinary}\\
t^{(S_i)}(\lambda)&=&\tr_a{\left[ {\cal L}_{a, L}^{(S_i,S_2)}(\lambda){\cal
L}_{a, L-1}^{(S_i,S_1)}(\lambda)\dots {\cal L}_{a, 2}^{(S_i,S_2)}(\lambda)
{\cal L}_{a, 1}^{(S_i,S_1)}(\lambda)\right]},
\label{transfer}
\ear
where we have assumed that $L$ is an even integer number.

The operator ${\cal L}_{a,j}^{(S_1,S_2)}(\lambda)$ is a rational solution to
the Yang-Baxter equation and is obtained by for instance the fusion
process \cite{KULISH}. It is given by
\eq
{\cal L}_{1,2}^{(S_1,S_2)}(\lambda)= \sum_{l=|S_1-S_2|}^{S_1+S_2}
f_{l}(\lambda) \check{P}_{l},
\label{Loperator}
\en
where $ f_{l}(\lambda)=\prod_{j=l+1}^{S_1+S_2}
\left(\frac{\lambda- j}{\lambda+  j}\right)
\prod_{j=1}^{2\min(S_{1},S_2)}(\lambda+|S_2 -S_1| +j)$ and $\check{P}_l$ is the
usual $SU(2)_l$ projector
\eq
\check{P}_{l}=\prod_{\stackrel{k=|S_1-S_2|}{k \neq l}}^{S_1+S_2}
\frac{\vec{S}_{1}\otimes \vec{S}_{2}-x_{k}}{x_{l}-x_{k}},
\en
with $x_{l}=\frac{1}{2}\left[l(l+1)-S_1(S_1+1)-S_2(S_2+1)\right]$ and the
$SU(2)$ generators
$\vec{S}_{a}=(\hat{S}_{a}^{x},\hat{S}_{a}^{y},\hat{S}_{a}^{z})$ for $a=1,2$.
The operator (\ref{Loperator}) has the following symmetry properties,
\begin{align}
  \mbox{Unitarity: } & {\cal L}_{12}^{(S_1,S_2)}(\lambda) {\cal
  L}_{12}^{(S_1,S_2)}(-\lambda)= \zeta_{S_1,S_2}(\lambda) 
\mbox{Id}_1 \otimes \mbox{Id}_2, \label{uni} \\
\mbox{Crossing:
  } & {\cal L}_{12}^{(S_1,S_2)}(\lambda) = (-1)^{2S_1} M_1 {\cal
  L}_{12}^{(S_1,S_2)}(-\lambda-1)^{t_{2}}
  M_{1}^{-1}, \label{cross} \\ \mbox{Regularity: } & {\cal
  L}_{12}^{(S_1,S_1)}(0)= (2S_1)! P_{1,2}^{(S_1)},
\end{align}
where $\zeta_{S_1,S_2}(\lambda)=\prod_{j=1}^{2\min(S_1,S_2)} ((|S_2-S_1|+j)^2
-\lambda^2)$. The matrix $M$ is an anti-diagonal matrix
$M_{i,j}=-(-1)^{i}\delta_{i,2S_{1}+2-j}$ and $P_{1,2}$ is the permutation
operator. From now on, we assume that $S_2\geq S_1$ without loss of
generality.

The transfer matrix $T(\lambda)$ can conveniently be re-written as
\eq
T(\lambda)=\tr_{\vec{a}}{\left[ L_{\vec{a},\overrightarrow{\frac{L}{2}}
}(\lambda) \dots L_{\vec{a},\vec{1}}(\lambda) \right]},
\label{newtransfer}
\en
where $\vec{j}=( 2 j-1, 2 j)$, $\vec{a}=(a_1,a_2)$ and
\eq
 L_{\vec{a},\vec{b}}(\lambda)= \left( 
{\cal L}_{a_1, b_2}^{(S_1,S_2)}(\lambda)  
{\cal L}_{a_2, b_2}^{(S_2,S_2)}(\lambda) \right) 
\left( {\cal L}_{a_1, b_1}^{(S_1,S_1)}(\lambda) 
{\cal L}_{a_2, b_1}^{(S_2,S_1)}(\lambda) \right).
\en

The conserved currents ${\cal J}^{(n)}$ are obtained by taking logarithmic
derivatives of the transfer matrix $T(\lambda)$
\eq
{\cal J}^{(n)}=\frac{\partial^n}{\partial\lambda^n} \ln{T(\lambda)} \Big|_{\lambda=0},
\en
evaluated at the regular point $L_{\vec{a},\vec{b}}(0)= C_0
P_{\vec{a},\vec{b}}$, where $P_{\vec{a},\vec{b}}=P_{a_2,b_2}^{(S_2)}
P_{a_1,b_1}^{(S_1)}$ and $C_0=(2S_1)!(2S_2)!\prod_{j=1}^{2S_1}(S_2-S_1+j)^2$.

The quantum Hamiltonian of the alternating spin chain is the first non-trivial
conserved current ${\cal H}={\cal J}^{(1)}$, which results in
\eq
{\cal H}=\sum_{k=1}^{L/2} h_{\vec{k},\overrightarrow{k+1} }, ~ \mbox{where} ~  h_{\vec{k},\overrightarrow{k+1}}=\frac{1}{C_0} P_{\vec{k},\overrightarrow{k+1}}\frac{\partial}{\partial\lambda} L_{\vec{k},\overrightarrow{k+1}}(\lambda)\Big|_{\lambda=0},
\en
where periodic boundary conditions are assumed. For illustration, the
Hamiltonian for case $S_1=1/2$, $S_2=1$ is given explicitly by \cite{DEVEGA}
\bear
{\cal H}^{(\frac{1}{2},1)}&=&\frac{2}{9}
\sum_{\text{even }i}
\left(
4\left\{\vec{\sigma}_{i-1}\cdot \vec{S}_{i}+\frac{5}{4}~ ;~\vec{S}_{i} \cdot
    \vec{\sigma}_{i+1}+\frac{3}{4}   \right\}-7\vec{\sigma}_{i-1}
\cdot \vec{\sigma}_{i+1}-\frac{1}{2}
\right.\nonumber\\
&&\left.
-\frac{3}{2}
\left\{
\vec{S}_{i}\cdot \vec{S}_{i+2}~;~
\vec{S}_{i} \cdot \vec{\sigma}_{i+1}+\vec{\sigma}_{i-1}\cdot\vec{S}_{i}
- \frac{3}{4} - \frac{1}{4}  \vec{S}_{i}\cdot \vec{S}_{i+2} 
\right\} \right).
\label{hamilt1o2S}
\ear

The exact computation of the thermal conductivity and its Drude weight as a
function of temperature is possible thanks to the fact that one can identify
the thermal current operator with one of the conserved currents underlying the
integrability of the model.

In order to establish the desired connection, we consider the local
conservation of energy in terms of a continuity equation. This relates the
time derivative of the local Hamiltonian $h_{\vec{k},\overrightarrow{k+1}}$ to
the divergence of the thermal current $j^{E}$, $\dot{h}=-\nabla j^{E}$.

As the time derivative leads to the commutator with the Hamiltonian, we obtain
\eq
\dot{h}_{\vec{k},\overrightarrow{k+1}}=\im 
\left[{\cal H}, h_{\vec{k},\overrightarrow{k+1}}(t) \right]=
-\im \left(j_{\overrightarrow{k+1}}^E(t) - j_{\vec{k}}^E(t) \right),
\en
where the local energy current $j_{\vec k}^E$ is given by
\eq
j_{\vec k}^E=\im \left[ h_{\overrightarrow{k-1},\vec k}, h_{\vec k,\overrightarrow{k+1}} \right],
\en
and the total thermal current is ${\cal J}_E=\sum_{k=1}^{L/2} j_{\vec k}^{E}$.

On the other hand, by inspection we find that the expression for ${\cal J}_E$
and the second logarithmic derivative of the transfer matrix ${\cal J}^{(2)}$
are closely related
\eq
{\cal J}_E={\cal J}^{(2)} + \im \frac{L}{2}\frac{\partial^2
{\Xi}(\lambda)}{\partial \lambda^2} \Big|_{\lambda=0},
\en
where $\Xi(\lambda)=\prod_{1\le i,j \le
2} \zeta_{S_i,S_j}(\lambda)$. Therefore the thermal current can be identified as 
one of the conserved currents underlying the integrability. This property is exploited for studying the thermal Drude
weight of the $(S_1,S_2)$ alternating spin chain in the thermodynamical limit.

We would like to stress that this identification is possible in the case of
rotational invariant vertex model discussed here. In this case we have the
staggering of spins $S_1$ and $S_2$ in the vertical and horizontal directions
of the classical vertex model. However, a similar identification for the
non-rotationally invariant case discussed on \cite{RIBEIRO} is still an open
question.

\section{Thermal Drude weight}
\label{THERMAL}

The transport coefficients are determined from the Kubo formula \cite{KUBO} in
terms of the expectation value of the thermal current ${\cal J}_E$, such that
\cite{ZOTOS,SAKAI}
\eq
D_{th}(T)=\beta^2 \left\langle {\cal J}_{E}^2 \right\rangle, ~~ \beta:=\frac{1}{T}.
\en

In order to calculate the expectation value $\left\langle {\cal J}_{E}^2
\right\rangle$, we introduce an auxiliary partition function $\bar{Z}$,
\eq
\bar{Z}=\tr{\left[\exp{\left(-\beta {\cal H}-\lambda_n {\cal
          J}^{(n)}\right)}\right]}.
\en
In this way, we obtain the expectation values of ${\cal J}^{(2)}$ through the
logarithmic derivative of $\bar{Z}$,
\eq
\left(\frac{\partial}{\partial \lambda_2}\right)^2
\llg{\bar{Z}}\Big|_{\lambda_2=0}=\left\langle {\cal J}_{E}^2 \right\rangle,
\en
where we used the fact that the expectation value of the thermal current in
thermodynamical equilibrium is zero $\left\langle {\cal J}_{E}
\right\rangle=0$.

To compute the partition function $\bar{Z}$, we follow the procedure
developed in \cite{SAKAI}. We rewrite the partition function $\bar{Z}$ in
terms of the row-to-row transfer matrix such that
\bear
\bar{Z}&=&\lim_{N\rightarrow \infty}\tr{\left[ T(u_1)\dots T(u_N)
        T(0)^{-N} \right]}, \nonumber\\
&=&\tr{\left[\exp{\left( \lim_{N\rightarrow \infty} \sum_{l=1}^{N}\{
        \llg{T(u_l)}- \llg{T(0)} \} \right)}\right]}.
\ear
The numbers $u_1,\dots,u_N$ are chosen in such a way that the following
relation is satisfied, 
\eq
\lim_{N\rightarrow \infty}\sum_{l=1}^{N}\{ \llg{T(u_l)}- \llg{T(0)} \}=-\beta
\frac{\partial}{\partial x}\llg{T(x)}\Big|_{x=0}+\lambda_n
\im^{n-1}\frac{\partial^n}{\partial x^n}\llg{T(x)}\Big|_{x=0}.
\en

In what follows we shall extend the results of \cite{RIBEIRO} in order to
include rotationally invariant classical vertex model associated with the
alternating spin chain. In doing so, we need to introduce quantum transfer
matrices $T^{(S_1,\{S_1,S_2\})}$ and $T^{(S_2,\{S_1,S_2\})}$ for spins $S_1$
and $S_2$. In this way, the partition
function $\bar{Z}$ in the thermodynamical limit can be written in terms of the
largest eigenvalues of the above mentioned quantum transfer matrices,
\eq
\lim_{L\rightarrow \infty}\frac{1}{L} \llg{\bar{Z}}=
\frac{1}{2}\llg{\Lambda^{(S_1)}(0)}+\frac{1}{2}\llg{\Lambda^{(S_2)}(0)},
\en
with
\bear
\llg{\Lambda^{(S_1)}(0)}&=&(-\beta + \lambda_n \frac{\partial^{n-1}}{\partial
x^{n-1}} ){\cal E}_1(x)\Big|_{x=0} + \left(K*\llg{Y^{(S_1)}}\right)(0), \\
\llg{\Lambda^{(S_2)}(0)}&=&(-\beta + \lambda_n \frac{\partial^{n-1}}{\partial
x^{n-1}} ){\cal E}_2(x)\Big|_{x=0} + \left(K*\llg{B\bar{B}}\right)(0),
\label{eigvalues}
\ear
where $K(x)=\frac{\pi}{\cosh{\left[ \pi x \right]}}$ and the functions ${\cal
E}_i(x)$ are given by ${\cal
E}_i(x)$\break\hfil
$= \int_{-\infty}^{\infty}\frac{\hat{\gamma}_i(k) e^{\im
kx}}{2\cosh{\left[k/2\right]}} dk$,
\bear
\hat{\gamma}_1(k)&=&\frac{e^{-|k|(S_1+\frac{1}{2})}\sinh[|k|(S_1-\frac{1}{2})]
+e^{-|k|(S_2-\frac{1}{2})} \sinh[|k|(S_1+\frac{1}{2})]}{\sinh[|k|/2]}, \\
\hat{\gamma}_2(k)&=&e^{-|k|S_2}\frac{\sinh{\left[|k| S_1 \right]}
+\sinh{\left[|k|S_2 \right]}}{\sinh{\left[|k|/2 \right]} }.
\ear
The auxiliary functions $Y^{(S_1)}(x)$, $B(x)$ and $\bar{B}(x)$ are required
to satisfy a closed set of functional equations. The procedure of deriving
from these equations a set of integral equations is similar to that described
in Ref. \cite{SUZUKI}, therefore we just present the final result
\eq
\left(
\begin{array}{c}
\llg{y^{(\frac{1}{2})}(x)} \\
\vdots \\
\llg{y^{(S_1)}(x)} \\
\vdots \\
\llg{y^{(S_2-\frac{1}{2})}(x)} \\
\llg{b(x)} \\
\llg{\bar{b}(x)}
\end{array}\right)=
\left(\begin{array}{c}
0 \\
\vdots \\
(-\beta+ \lambda_n \frac{\partial^{n-1}}{\partial x^{n-1}} ) d(x)  \\
\vdots \\
0 \\
(-\beta+ \lambda_n \frac{\partial^{n-1}}{\partial x^{n-1}} ) d(x)  \\
(-\beta+ \lambda_n \frac{\partial^{n-1}}{\partial x^{n-1}} ) d(x)  
\end{array}\right)
+
{\cal K}*
\left(\begin{array}{c}
\llg{Y^{(\frac{1}{2})}(x)} \\
\vdots \\
\llg{Y^{(S_1)}(x)} \\
\vdots \\
\llg{Y^{(S_2-\frac{1}{2})}(x)} \\
\llg{B(x)} \\
\llg{\bar{B}(x)}
\end{array}\right),
\label{nlieqnew}
\en
where $d(x)=\frac{\pi}{\cosh{\left[ \pi x \right]}}$ and the symbol $*$
denotes the convolution $f*g(x)=\int_{-\infty}^{\infty} f(x-y)g(y)\frac{dy}{2\pi}$.

The kernel matrix reads explicitly
\eq
{\cal K}(x)= 
\left(
\begin{array}{cccccccc}
0 & K(x) & 0 & \cdots & 0 & 0 & 0 & 0 \\
K(x) & 0 &  K(x) &   & \vdots & \vdots & \vdots & \vdots  \\
 0 & K(x) & 0  &  &  & 0 & 0 & 0 \\
\vdots &  &   &   & 0 & K(x) & 0 & 0 \\
0 & 0 & \cdots  & 0 & K(x) & 0 & K(x) & K(x) \\
0 & 0 & \cdots  & 0 & 0 & K(x) & F(x) & -F(x+\im) \\
0 & 0 & \cdots  & 0 & 0 & K(x) & -F(x-\im) & F(x) \\
\end{array}\right),
\label{Kernel-x}
\en
where $F(x)=\int_{-\infty}^{\infty}\frac{e^{-|k|/2+\im k
x}}{2 \cosh{\left[k/2 \right]}} dk $ and $K(x)$ was defined in Eq.(\ref{eigvalues}).

\begin{figure}[h]
\begin{center}
\includegraphics[height=10cm,width=10cm]{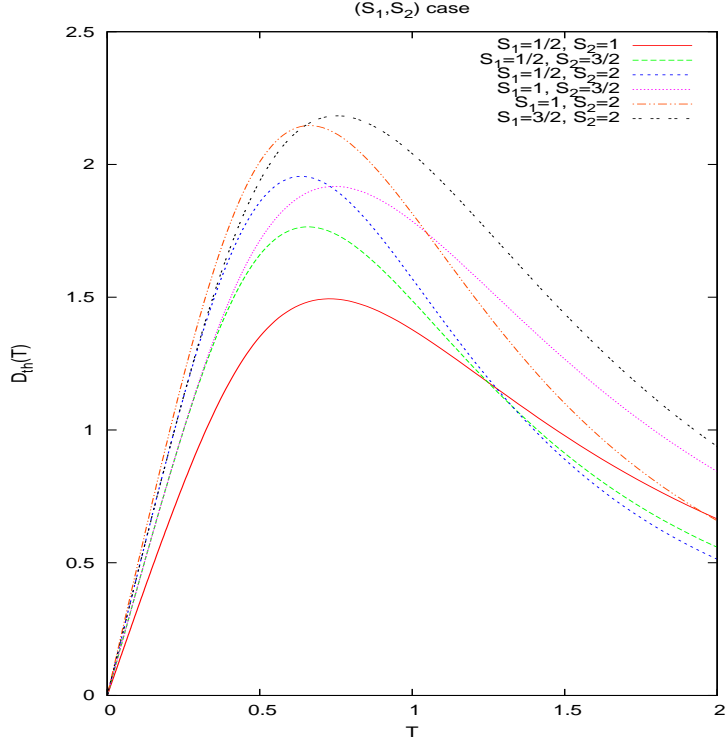}
\caption{Thermal Drude weight $D_{th}(T)$ as function of temperature for
  $(S_1=1/2,S_2=1), (S_1=1/2,S_2=3/2), (S_1=1/2,S_2=2), (S_1=1,S_2=3/2),
 (S_1=1,S_2=2)$ and $(S_1=3/2,S_2=2)$.}
\label{pic4}
\end{center}
\end{figure}
These equations have a different driving term structure in comparison with
those for the homogeneous case discussed in \cite{RIBEIRO}. Therefore, the
equations (\ref{nlieqnew}) constitute a new set of equations suitable for the
description of the thermal Drude weight of the alternating spin chains.

Finally, the thermal Drude weight can be written as
\eq
D_{th}(T)=\beta^2 \left\langle {{\cal J}^{(2)}}^2 \right\rangle=\frac{1}{2}\beta^2
\left(\frac{\partial}{\partial\lambda_2}\right)^2 \left[\llg{\Lambda}^{(S_1)}(0) 
+\llg{\Lambda}^{(S_2)}(0)\right]\Big|_{\lambda_2=0}.
\en

In Figure \ref{pic4}, we show numerical results for the thermal Drude weight
as a function of temperature for specific choices of the spins $(S_1,S_2)$. At
low temperatures the data exhibit a linear temperature dependence. The
low-temperature asymptotics is accessible by an analytical treatment of the
non-linear integral equations similar to \cite{SUZUKI}. From this we find that the
thermal Drude weight $D_{th}(T)$ is generally a linear function of temperature
and is proportional to the central charge of the system $c
= \frac{3S_1}{S_1+1} + \frac{3(S_2-S_1)}{(S_2-S_1)+1}$\cite{ALADIM}. Its explicit expression reads
\eq
D_{th}(T)\simeq\frac{\pi v_s c}{12} T,
\label{Dweight}
\en
where $v_s=2 \pi$ is the sound velocity. These results are in
agreement with those of the special case of homogeneous spin chains
($S_1=S_2=S$) \cite{RIBEIRO}.

The low-temperature asymptotics (\ref{Dweight}) is remarkable. This result may
be derived on the hypothetical grounds of a particle picture of the thermal
transport. If we assume that the elementary excitations of the system are the
carriers of the thermal transport, we would expect that the thermal
conductivity $\kappa$ of a system with finite mean free path $l$ (due to
imperfections of the lattice) is $\kappa=v_s ~ c(T) ~ l$ where $v_s$ is the
velocity of the elementary excitations and $c(T)$ is the specific heat. On the
other hand, the existence of a finite mean free path defines a mean life
time $\tau$ over which the Drude weight broadens to give a finite conductivity
$\kappa=D_{th} ~ \tau$. The condition that both formulas agree is
$D_{th}=v_s^2 ~ c(T)$ which is --up to a numerical factor of order
1-- equivalent to (\ref{Dweight}). It is remarkable
that this reasoning yields the correct result despite the insufficient
grounds: For generic Tomonaga-Luttinger liquids and conformally invariant
field theories there are no delta-function peaks in the spectral function of the
single-particle Green's functions.

\section{Conclusion}
\label{CONCLUSION}
In this paper we have obtained a set of non-linear integral equations allowing
for the explicit calculation of the thermal Drude weight for the integrable
alternating ($S_1,S_2$) spin chains. We have solved the equations numerically
for specific values of $S_1$ and $S_2$. At low-temperatures we observe linear
temperature dependence of the thermal Drude weight. This linear behavior is
confirmed by the analytical low-temperature asymptotic solution showing that the
coefficient of proportionality is the central charge of the
system $c = \frac{3S_1}{S_1+1} + \frac{3(S_2-S_1)}{(S_2-S_1)+1}$. 
The existence of a non-zero Drude weight at finite temperature is a
signature of anomalous thermal transport \cite{ZOTOS}.
We hope our findings will be useful for experimental investigations of transport properties of quasi
one-dimensional mixed spin systems, e.g \cite{EXP4}.

Finally, we would like to remark that our results may be further generalized
by including the alternation of an arbitrary number of different spins $S_i,
i=1,\dots,M$. Another direction of generalization would be the extension to
anisotropic spin chains.

\section*{Acknowledgments}
The authors thank the Volkswagen Foundation for financial
support. G.A.P. Ribeiro also thanks FAPESP for financial support.

\end{document}